\documentclass[manuscript]{acmart}

\usepackage[ruled,lined]{algorithm2e}
\usepackage{xspace}
\usepackage{subfig}
\usepackage[dvipsnames]{xcolor}
\usepackage{tikz}
\usepackage{flushend}
\usepackage{soul}

\AtBeginDocument{%
  }

\graphicspath{{figs/}}

\ifx\figurename\undefined \def\figurename{Figure}\fi
\renewcommand{\figurename}{Fig.}
\newcommand{\para}[1]{\textit{\textbf{#1}} }
\renewcommand{\subparagraph}[1]{\underline{\textit{#1}} }

\newcommand{\Sect}[1]{Sec.~\ref{#1}}
\newcommand{\Fig}[1]{Fig.~\ref{#1}}

\def\cG{{\mathcal{G}}}
\def\cL{{\mathcal{L}}}
\def\cI{{\mathcal{I}}}

\def\cT{{\mathcal{T}}}

\newcommand{\mode}[1]{\underline{\textsc{#1}}\xspace}

\newcommand{\proj}{\textsc{ByteSplat}\xspace}

\definecolor{myorange}{RGB}{255, 212, 121}

\newcommand{\PBox}[1]{\vspace*{.05cm}\noindent\fbox{\parbox{.98\columnwidth}{\vspace*{.05cm}{#1}}}\vspace*{.05cm}}

\begin{document}

\title[\proj: Efficient Distributed 3D Gaussian Splatting Training]{\proj: Efficient Distributed 3D Gaussian Splatting Training via Intra- and Inter-GPU communication reduction}






\author{Shuo Wu}
\orcid{}
\email{wushuo123@sjtu.edu.cn}

\author{He Zhu}
\orcid{0009-0006-0835-7153}
\email{zhcon16@sjtu.edu.cn}

\author{Han Zhao}
\orcid{}
\affiliation{%
  \institution{Shanghai Jiao Tong University}
  \city{Shanghai}
  \country{China}
}
\email{zhao-han@cs.sjtu.edu.cn}

\author{Xiaohui Zhang}
\orcid{}
\email{zhangxiaohui06@ieisystem.com}

\author{Yaqian Zhao}
\orcid{}
\email{zhaoyaqian@ieisystem.com}

\author{Hui Wei}
\orcid{}
\affiliation{%
  \institution{IEIT SYSTEMS Co., Ltd.}
  \city{Beijing}
  \country{China}
}
\email{weihui@ieisystem.com}

\author{Ruyang Li}
\orcid{}
\email{liruyang@ieisystem.com}

\author{Hongzhi Shi}
\orcid{}
\email{shihzh@ieisystem.com}

\author{Lihua Lu}
\orcid{}
\affiliation{%
  \institution{IEIT SYSTEMS Co., Ltd.}
  \city{Beijing}
  \country{China}
}
\email{lulihua@ieisystem.com}

\author{Jingwen Leng}
\orcid{0000-0002-5660-5493}
\email{leng-jw@sjtu.edu.cn}

\author{Yu Feng}
\authornote{Corresponding Authors.}
\orcid{0000-0002-2192-5737}
\email{y-feng@sjtu.edu.cn}

\author{Minyi Guo}
\authornotemark[1]
\orcid{0000-0003-0034-2302}
\affiliation{%
  \institution{Shanghai Jiao Tong University}
  \city{Shanghai}
  \country{China}
}
\email{guo-my@sjtu.edu.cn}

\renewcommand{\shortauthors}{Shuo Wu et al.}

\begin{abstract}

3D Gaussian Splatting (3DGS) enables photorealistic scene reconstruction, but training large-scale scenes requires substantial memory and computation.
Distributing training across multiple GPUs increases available memory capacity, yet its performance is strictly constrained by data movement.
We identify two dominant bottlenecks: repeated off-chip memory accesses during forward and backward rasterization, and inter-GPU communication of partial Gaussian gradients.
To address these bottlenecks, we present \proj, a distributed 3DGS training framework that jointly reduces intra- and inter-GPU data movement.
First, \proj fuses forward rasterization and backward rasterization into a single GPU kernel, retaining intermediate results on-chip to eliminate redundant off-chip transfers.
Second, to alleviate the increased on-chip storage by fused rasterization, \proj introduces hardware-aware pruning that considers both rendering quality and per-tile shared-memory constraints, increasing the overall performance of fused execution.
Lastly, \proj exploits gradient sparsity to eliminate zero partial gradients from inter-GPU communication.
GPU-efficient encoder and decoder kernels compact the remaining records and directly aggregate the received gradients on their owner GPUs, reducing both communication volume and local aggregation overhead.
We evaluate \proj across six datasets.
Compared with the baseline, \proj reduces intra-GPU off-chip traffic and backward inter-GPU communication volume by 63.4\% and 65.8\%, respectively.
On eight GPUs, \proj achieves up to 6.1$\times$ training speedup while preserving reconstruction quality.

\end{abstract}

  


\maketitle

\section{Introduction}
\label{sec:intro}

3D Gaussian Splatting (3DGS)~\cite{kerbl20233d, fang2024mini, mallick2024taming, fan2023lightgaussian, zhu2026seele} has emerged as a promising representation for photorealistic scene reconstruction. 
More recently, 3DGS has also become the foundational substrate for world models, including Spark~\cite{spark} and HY-World~2.0~\cite{hyworld2026}. 
However, scaling 3DGS training to large scenes~\cite{kerbl2024hierarchical, wu2025blockgaussian, liu2024citygaussian, ren2024octree} requires optimizing an increasingly large number of Gaussians. 
Studies~\cite{zhao2025on} show that training these large-scale scenes can require more than 80~GB of memory, far exceeding the capacity of a single GPU.
Therefore, distributing Gaussians across multiple GPUs is necessary to provide sufficient memory capacity.

In mainstream distributed 3DGS training frameworks~\cite{zhao2025on, jia2026splaxel}, Gaussians are typically partitioned across multiple GPUs. 
At the beginning of each training iteration, each GPU retrieves the Gaussians required to render its assigned image regions from other GPUs. 
After completing its local computation, each GPU then sends partial Gaussian gradients back to the corresponding owner GPUs for gradient aggregation and parameter updates. 
Our characterization in \Sect{sec:motiv} shows that this computing paradigm leads to major performance bottlenecks.
Thus, it is important for us to revisit its training dataflow.

\para{Bottlenecks.}
In \Sect{sec:motiv}, we identify two dominant bottlenecks during 3DGS training.
The first bottleneck is the data movement within local GPU computation.
Specifically, forward and backward rasterization introduce substantial off-chip memory traffic. 
During tile-based forward rasterization, each thread block loads the related Gaussian attributes from off-chip DRAM into on-chip SRAMs.
Later on, backward rasterization reloads the same attributes again and writes a large volume of partial Gaussian gradients to off-chip memory for aggregation.
Consequently, forward and backward rasterization account for 29.9\% of the end-to-end training time and 37.9\% of the off-chip memory traffic.

The second bottleneck is from the inter-GPU communication. 
Before rasterization, Gaussian attributes are transferred from the owner GPUs to the rendering GPUs responsible for rendering the corresponding image regions and computing partial gradients.
After backward rasterization, the partial gradients are then returned to the owner GPUs for aggregation and parameter updates.
We show that this gradient aggregation stage is one of the bottleneck stages and accounts for 23.9\% of the total training time.
These observations show that distributed 3DGS training is constrained by both intra-GPU off-chip traffic and inter-GPU gradient communication.
This motivates us to design \proj, which optimizes data movement at both levels.

\para{Fused Rasterization.}
To reduce intra-GPU data movement, we propose a mega-kernel that fuses forward rasterization, loss computation, and backward rasterization into a single mega-kernel.
Doing so, the kernel retains all necessary attributes on-chip throughout these stages and eliminates off-chip transfers of intermediate data between stages.
In the conventional pipeline, forward and backward rasterization execute as separate kernels and exchange data through off-chip memory.
Consequently, Gaussian attributes accessed during the forward rasterization must be reloaded during the backward rasterization, while the intermediate results between these two must be written back to DRAM.
In contrast, our mega-kernel retains Gaussian attributes and intermediate results on-chip across both stages.
For each tile, it loads the Gaussian data once, performs forward rasterization and the corresponding backward computation within a single kernel invocation.
This way, it only writes the final outputs required by subsequent stages to off-chip memory.
In \Sect{sec:eval:gpu}, we show that this fused dataflow reduces off-chip memory traffic by 63.4\%.

\para{Hardware-Aware Pruning.}
Although fused rasterization reduces off-chip memory traffic, a naive mega-kernel implementation can impose substantial pressure on on-chip memory.
Because fusing forward and backward rasterization extends the lifetime of Gaussian attributes and intermediate results, it increases the shared-memory requirement of each thread block.
This pressure can be further exacerbated by the uneven distribution of Gaussians across image tiles.
Those tiles whose working sets exceed the per-block shared-memory limit cannot use the fused kernel.
Even below this limit, large shared-memory allocations can reduce thread-block occupancy and offset the benefits of fusion.

To address this challenge, we propose a hardware-aware Gaussian pruning strategy that explicitly accounts for the on-chip resource constraints of the target GPU.
During training, our strategy first identifies image tiles for which fused execution is infeasible without pruning.
For overloaded tiles, we then progressively remove low-importance Gaussians associated with these tiles and increase the fraction of tiles that satisfy the mega-kernel's register and shared-memory budgets.
Unlike conventional pruning methods that primarily target model size, our method incorporates hardware resource constraints and directly optimizes the execution efficiency of fused rasterization.
By integrating pruning into the training process, the remaining Gaussians can adapt to compensate for the removed primitives.
\proj can alleviate on-chip resource pressure and improve fused-kernel efficiency while preserving reconstruction quality.

\para{Zero-Eliding Gradient Communication.}
To reduce inter-GPU data movement, we exploit the sparsity of partial Gaussian gradients. 
Existing distributed systems return the partial Gaussian gradients of every Gaussian received for local rendering, regardless of whether that gradient contributes to the final parameter update.
Our characterization shows that up to 72.0\% of these partial gradients are zero.
Communicating zero gradients consumes interconnect bandwidth without contributing to gradient aggregation.

Based on this observation, we propose zero-eliding gradient communication, which transfers only nonzero partial gradients required for gradient aggregation.
Before communication, \proj identifies partial gradients whose components are all exactly zero and compacts the remaining gradient results.
After receiving the compacted data, each owner GPU decodes the representation and accumulates the nonzero gradients into the corresponding Gaussian parameters.
To make compaction and reconstruction efficient, we further develop specialized encoder and decoder kernels that perform encoding and decoding with low overhead.
Together, these techniques reduce inter-GPU communication volume by 46.2\% and accelerate the end-to-end training time by 1.2$\times$.

\para{Result.}
We integrate these three techniques into \proj, an efficient distributed 3DGS training framework.
We evaluate \proj across six datasets with up to 21.6 million Gaussians.
Compared with other baselines, \proj reduces intra-GPU and backward inter-GPU data traffic by 63.4\% and 65.8\%, respectively.
Together, \proj achieves up to 6.1$\times$ speedup, while retaining reconstruction quality.
The contributions of this paper are as follows:

\begin{itemize}
    \item We characterize the intra- and inter-GPU data movement in distributed 3DGS training and identify that data movement is the main performance bottleneck.
    \item We propose fused rasterization, which fuses forward rasterization, loss computation, and backward rasterization, reducing off-chip transfers of intermediate results.
    \item We propose hardware-aware training to maximize the performance benefits of the fused rasterization.
    \item We introduce zero-eliding gradient communication, which avoids transmitting zero-valued partial gradients between rendering and owner GPUs with low processing overhead.
    \item We integrate these techniques into \proj and achieve up to 6.1$\times$ speedup over 8-GPU baseline.
\end{itemize}
\section{Background}
\label{sec:bg}

We first describe the single-GPU pipeline of 3DGS training in \Sect{sec:bg:pipeline}.
We then explain how distributed systems partition the training workload across multiple GPUs in \Sect{sec:bg:distributed}.

\begin{figure*}[t]
    \centering
    \includegraphics[width=\textwidth]{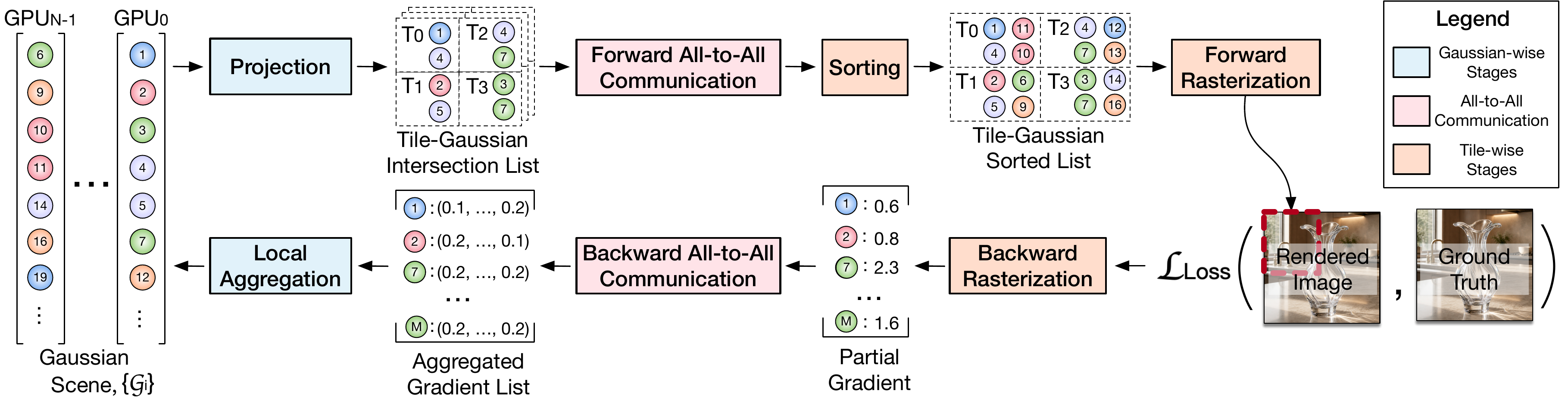}
    \caption{
    Overview of the distributed 3DGS training pipeline.
    Each owner GPU projects its locally stored Gaussians for each selected camera view and performs forward Gaussian all-to-all communication to send the attributes of intersected Gaussians to the rendering GPUs responsible for the intersected image tiles.
    The rendering GPUs then construct and sort tile-Gaussian lists, perform forward and backward rasterization, and compute partial Gaussian gradients.
    Backward gradient all-to-all communication then returns these partial gradients to the corresponding owner GPUs for local aggregation and parameter updates.
    Blue, pink, and orange boxes denote Gaussian-wise stages, inter-GPU communication, and tile-wise stages, respectively.}
    \label{fig:pipeline}
\end{figure*}

\subsection{Single-GPU Training Pipeline}
\label{sec:bg:pipeline}

The goal of 3DGS training is to optimize a set of Gaussians, $\mathcal{G}=\{G_i\}$, that represent a 3D scene.
Each Gaussian $G_i$ contains 3D information, opacity, and a set of spherical-harmonic coefficients that describe its view-dependent color.
Given an initial point cloud and a set of posed training images, 3DGS training iteratively refines its Gaussian representation via a \textit{forward pass} and a \textit{backward pass}~\cite{kerbl20233d}.
The overall pipeline is shown in \Fig{fig:pipeline}.

\para{Forward Pass.}
The forward pass renders an image from a selected camera view and computes its loss against the corresponding ground-truth image with four stages: \textit{projection}, \textit{sorting}, \textit{forward rasterization}, and \textit{loss computation}.

\subparagraph{Projection.}
Given a camera pose, projection first transforms each Gaussian from the world coordinate system into the camera coordinate system.
It then projects the Gaussian onto the 2D image plane.
This stage operates Gaussian-by-Gaussian.
Based on the projected results, this stage determines which image tiles are intersected by each Gaussian and generates a tile-Gaussian intersection list.

\subparagraph{Sorting.}
After projection, each image tile obtains a list of intersecting Gaussians.
These Gaussians are then sorted by their depths to the camera, from the nearest to the farthest.

\subparagraph{Forward Rasterization.}
This stage renders the image tile by tile.
Each pixel traverses the Gaussians in its tile-Gaussian sorted list and performs $\alpha$-checking to determine whether a Gaussian contributes to that pixel.
For a pixel $\mathbf{p}$, its color is computed through front-to-back alpha compositing:
\begin{equation}
    \mathbf{C}(\mathbf{p})
    =
    \sum_{i=1}^{N_{\mathbf{p}}}
    T_i \alpha_i \mathbf{c}_i,
    \qquad
    T_i
    =
    \prod_{j=1}^{i-1}(1-\alpha_j),
    \label{eq:alpha-compositing}
\end{equation}
where $N_{\mathbf{p}}$ is the number of Gaussians traversed by pixel $\mathbf{p}$.
Here, $\alpha_i$ denotes the effective opacity of the $i$th Gaussian at pixel $\mathbf{p}$, and $\mathbf{c}_i$ denotes its view-dependent RGB color.
$T_i$ represents the accumulated transmittance before that Gaussian.
Rasterization terminates once the transmittance $T_i$ falls below a threshold, $\theta_{T}$.

\subparagraph{Loss Computation.}
After rendering the selected view, the pipeline compares the rendered image with its ground-truth image.
The training objective often combines an L1 loss $\cL_1$ with a structural similarity loss $\cL_\mathrm{SSIM}$:
\begin{equation}
\cL = (1-\lambda)\cL_{1} + \lambda\cL_{\mathrm{SSIM}},
\label{eq:training-loss}
\end{equation}
where $\lambda$ controls the relative weights of the two terms.

\para{Backward Pass.}
The backward pass propagates the pixel-wise loss to individual Gaussians.
It consists of \textit{backward rasterization}, \textit{gradient aggregation}, and \textit{parameter updates}.

\subparagraph{Backward Rasterization.}
This stage propagates the pixel-wise loss gradients to individual Gaussians.
Loss computation first computes the color gradient of each rendered pixel, $\partial\cL/\partial\mathbf{C}(\mathbf{p})$.
Then, backward rasterization performs computation tile-by-tile.
Specifically, it traverses the corresponding tile-Gaussian sorted list and computes the partial gradient contributed by each pixel-Gaussian pair:
\begin{equation}
    \mathbf{g}_{\mathbf{p},i}
    =
    \frac{\partial \cL}
         {\partial \mathbf{C}(\mathbf{p})}
    \frac{\partial \mathbf{C}(\mathbf{p})}
         {\partial \widetilde{G}_i},
\label{eq:partial-gaussian-gradient}
\end{equation}
where $\widetilde{G}_i$ denotes the screen-space attributes of Gaussian $G_i$.
Since one Gaussian can contribute to multiple pixels, backward rasterization generates multiple partial gradients for one Gaussian, which need to be aggregated later.

\subparagraph{Gradient Aggregation and Update.}
We perform Gaussian-wise gradient updates.
For a Gaussian $G_i$, its gradient is the sum of the partial gradients from all pixels to which it contributes:
\begin{equation}
\mathbf{g}_i =
    \sum_{\mathbf{p}:i\in\cI(\mathbf{p})}
    \frac{\partial \cL_{\mathbf{p}}}
         {\partial \widetilde{G}_i},
\label{eq:gradient-aggregation}
\end{equation}
where $\cI(\mathbf{p})$ is the set of Gaussians contributing to pixel $\mathbf{p}$, $\cL_{\mathbf{p}}$ is the pixel-wise loss of that pixel, and $\widetilde{G}_i$ denotes the screen-space attributes of Gaussian $G_i$.
Because one Gaussian can contribute to many pixels, current GPU implementations commonly rely on atomic operations to accumulate these partial gradients correctly.
After aggregation, the gradients are backpropagated through Gaussian projection and color evaluation to obtain gradients of the original 3D Gaussian parameters.
The optimizer then updates the Gaussian positions, scales, rotations, opacities, and SH coefficients.

\subsection{Distributed Training Pipeline}
\label{sec:bg:distributed}

To support large-scale scenes, distributed 3DGS training partitions both Gaussians and rendering workloads across multiple GPUs~\cite{zhao2025on}.
As shown in \Fig{fig:pipeline}, different stages expose different execution granularities.
Thus, distributed training adopts different data partitions for different stages.

\para{Overview.}
At initialization, all Gaussians are divided into subsets and distributed across the GPUs.
Each Gaussian is assigned to an \textit{owner GPU}, which stores its trainable parameters and optimization state.
This Gaussian-wise partition is used in projection and Gaussian update, where each GPU processes its locally owned Gaussians independently.

For the remaining stages, each selected training image is divided into fixed-size tiles, e.g., tiles of $16\times16$ pixels.
The tiles are partitioned into subsets.
Each subset is assigned to a \textit{rendering GPU}.
Each rendering GPU performs sorting, forward rasterization, loss computation, and backward rasterization for its assigned tiles.
Because different tiles may have different Gaussian workloads, the tile assignment can be periodically adjusted to mitigate workload imbalance.

Note that, the owner GPUs and the rendering GPUs are two logical roles of the same physical GPUs.
Because the Gaussians required for a tile may be distributed across multiple owner GPUs, a rendering GPU generally does not locally store all the Gaussians for its assigned tiles. 
Therefore, Gaussian data need to be exchanged in distributed training when transitioning between Gaussian-wise and tile-wise stages.
Next, the following paragraphs describe the two communication points introduced by these transitions.

\para{Forward Gaussian Communication.}
During this stage, for each selected camera view, each owner GPU projects its locally owned Gaussians and identifies the rendering GPUs responsible for their intersected tiles.
Based on their projected results, the owner GPU determines which image tiles each Gaussian intersects and identifies the rendering GPUs responsible for those tiles.
The owner GPUs then perform sparse all-to-all communication to send the required Gaussians to the corresponding rendering GPUs.
Here, sparsity refers to the subset of Gaussians exchanged between GPU pairs.
For each Gaussian, we transfer its Gaussian depth, transmittance, and color attributes required by the subsequent stages.
After receiving these attributes, each rendering GPU constructs the tile-Gaussian intersection lists and performs sorting, forward rasterization, and loss computation for its assigned tiles.
This communication transitions the pipeline from the Gaussian-wise execution to the tile-wise execution.

\para{Backward Gradient Communication.}
The second transition happens when each rendering GPU finishes backward rasterization and prepares to send the partial gradients back to the corresponding owner GPUs.
Because a Gaussian may contribute to tiles assigned to multiple rendering GPUs, its partial gradients may be distributed across several GPUs.
The rendering GPUs therefore perform reverse sparse all-to-all communication to send the partial gradients, together with their Gaussian IDs, back to the corresponding owner GPUs.
Each owner GPU aggregates the received partial gradients for its locally owned Gaussians and updates the Gaussian parameters.
In this case, the pipeline is transitioned from tile-wise execution to Gaussian-wise execution.
\section{Motivation}
\label{sec:motiv}

\begin{figure*}[t]
\centering
  \begin{minipage}[t]{0.49\textwidth}
    \centering
    \includegraphics[width=0.9\textwidth]{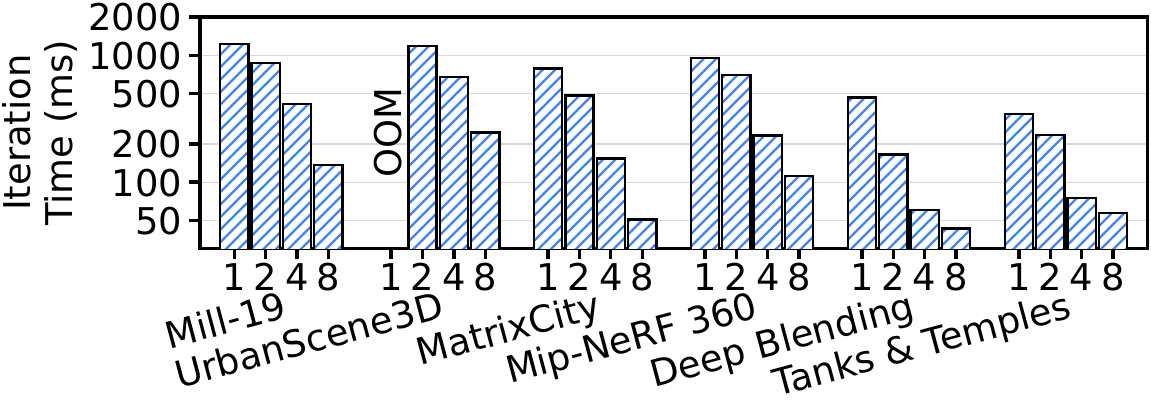}
    \caption{The Per-iteration Latency as the GPU Count Increases.}
    \label{fig:exe_latency}
  \end{minipage}
  \hfill
  \begin{minipage}[t]{0.49\textwidth}
    \centering
    \includegraphics[width=0.9\textwidth]{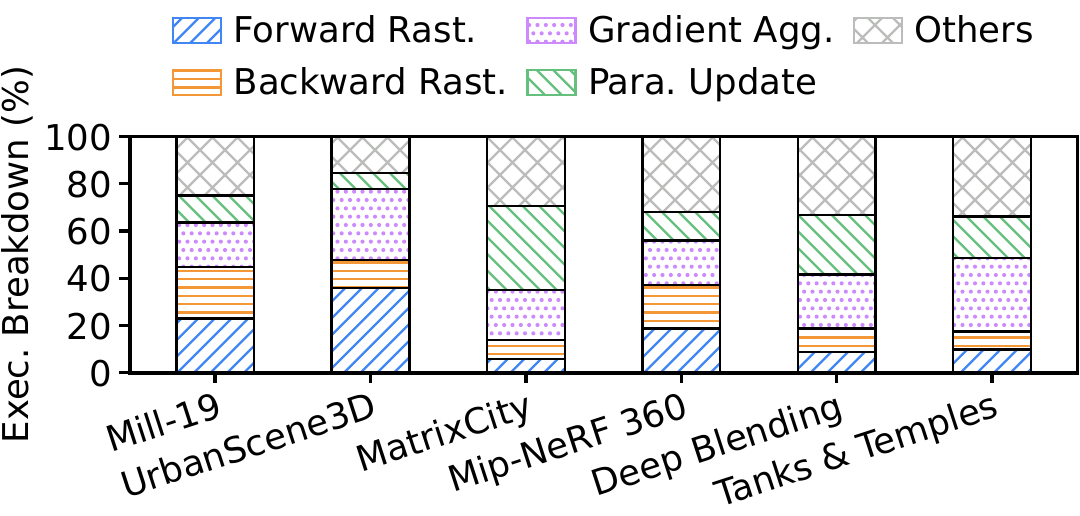}
    \caption{Execution-time Breakdown on A6000.}
    \label{fig:exec_breakdown}
  \end{minipage}
\end{figure*}

With the distributed training pipeline described in \Sect{sec:bg}, we characterize the performance of the original 3DGS training~\cite{zhao2025on} across six datasets~\cite{Knapitsch2017, hedman2018deep, barron2022mipnerf360, lin2022capturing, turki2022mega, li2023matrixcity} and show four key insights that motivate our \proj design.

\vspace{2pt}
\PBox{\textbf{Insight 1:} \textit{Forward/backward rasterization, gradient aggregation, and parameter update dominate training time.}}
\vspace{2pt}

First, \Fig{fig:exe_latency} shows the per-iteration latency of distributed 3DGS training as the GPU count increases from one to eight.
We profile training on six datasets, including three large-scale datasets~\cite{lin2022capturing, turki2022mega, li2023matrixcity} and three small-scale datasets~\cite{Knapitsch2017, hedman2018deep, barron2022mipnerf360}.
On a single GPU, iteration latency ranges from hundreds of milliseconds to over one second.
Although additional GPUs reduce this latency, the overall per-iteration latency is still significant.
For example, Mill-19~\cite{li2023matrixcity} requires approximately 1.2~seconds per iteration on one GPU and 0.15~seconds on eight GPUs.
These results motivate us to examine the execution-time breakdown and identify the bottleneck stages in the entire pipeline.

\Fig{fig:exec_breakdown} shows the execution breakdown of distributed training.
Here, we evaluate over the six datasets on eight Nvidia A6000 GPUs~\cite{a6000}.
Here, gradient aggregation includes the communication of partial gradients and their subsequent accumulation on the owner GPUs.
Across the evaluated scenes, forward rasterization, backward rasterization, distributed gradient aggregation, and the following parameter updates take 17.0\%, 13.0\%, 23.8\%, and 18.1\% of the training time, respectively.
Together, these four stages consume over 71.9\% of the total execution time.
Although their individual contributions vary across scenes, these stages collectively account for most training time.
Therefore, accelerating all these stages is critical to improving the end-to-end performance of distributed 3DGS training.

\begin{figure*}[t]
\centering
  \subfloat[Forward rasterization.]{
    \includegraphics[width=0.48\textwidth]{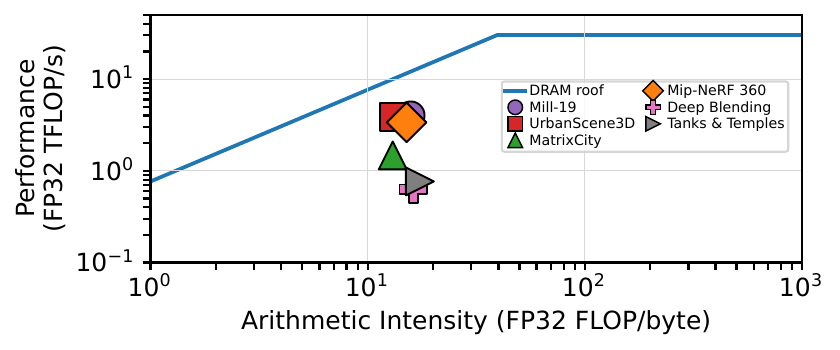}
    \label{fig:roofline_forward}
  }
  \hfill
  \subfloat[Backward rasterization.]{
    \includegraphics[width=0.48\textwidth]{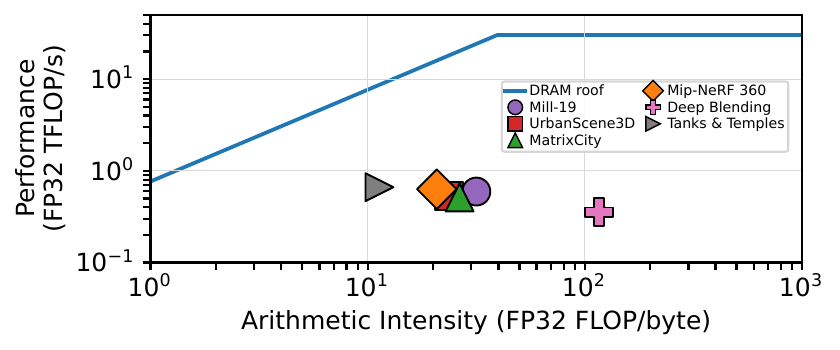}
    \label{fig:roofline_backward}
  }
  \caption{The Roofline Analysis of Forward and Backward Rasterization.}
  \label{fig:roofline}
\end{figure*}

\vspace{2pt}
\PBox{\textbf{Insight 2:} \textit{Forward and backward rasterization are memory-bound and incur redundant off-chip memory accesses.}}
\vspace{2pt}

To understand the root cause of the rasterization overhead, \Fig{fig:roofline} shows the roofline analysis of two bottleneck stages, forward rasterization and backward rasterization.
The numbers are measured on eight Nvidia A6000 GPUs. 
Overall, these two stages are primarily memory-bound.
This is because each thread block repeatedly loads the Gaussian attributes associated with an image tile from off-chip DRAM into on-chip memory in both forward and backward rasterization.
Overall, repeated Gaussian-attribute accesses and intermediate-data transfers account for 63.4\% of the off-chip traffic in these two stages.
These results show an opportunity to reduce rasterization latency by retaining reusable data on-chip across the forward and backward rasterization.

\begin{figure*}[t]
\centering
  \begin{minipage}[t]{0.49\textwidth}
    \centering
    \includegraphics[width=0.9\textwidth]{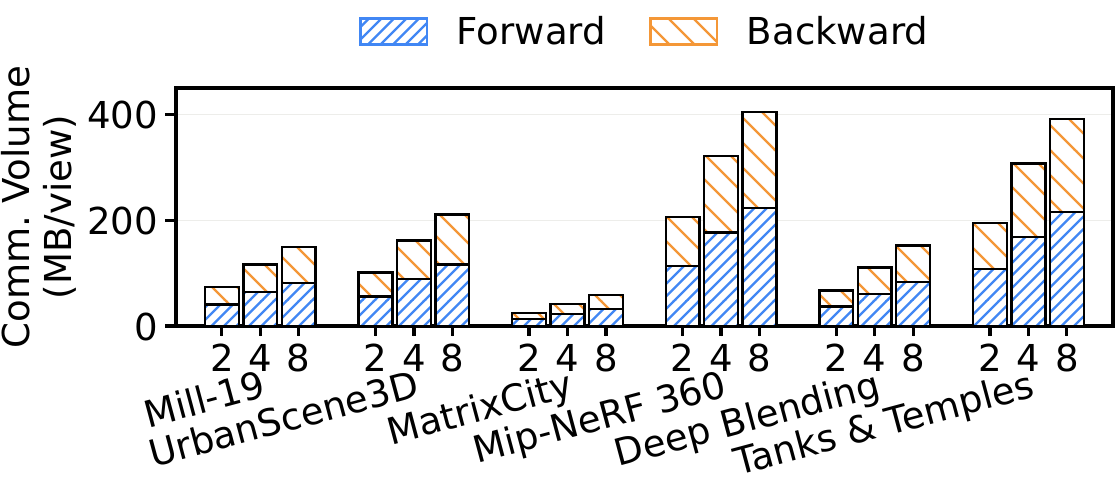}
    \caption{The Volume of Communication During Forward and Backward.}
    \label{fig:data_comm}
  \end{minipage}
  \hfill
  \begin{minipage}[t]{0.49\textwidth}
    \centering
    \includegraphics[width=0.9\textwidth]{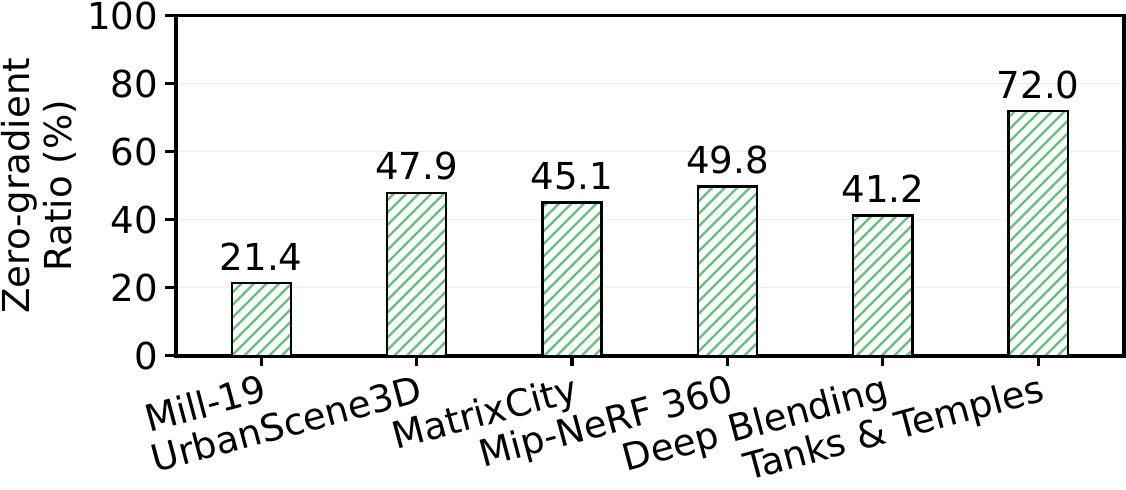}
    \caption{The Percentage of Zero Gradients During Backward Gradient Communication.}
    \label{fig:gradient}
  \end{minipage}
\end{figure*}

\vspace{2pt}
\PBox{\textbf{Insight 3:} \textit{Backward gradient communication is nontrivial in the overall inter-GPU data communication.}}
\vspace{2pt}

Distributed 3DGS training introduces inter-GPU communication at two places in the training pipeline: \textit{forward Gaussian communication} and \textit{backward gradient communication}.
Specifically, after projection, forward Gaussian communication transfers the projected Gaussian attributes from their owner GPUs to rendering GPUs.
After backward rasterization, backward gradient communication returns the Gaussian gradients from rendering GPUs to owner GPUs.

\Fig{fig:data_comm} compares the communication volumes of these two stages across six datasets.
Overall, as the GPU count increases from two to eight, the total all-to-all communication volume nearly doubles.
This growth increases with the number of GPUs because distributing image tiles across more rendering GPUs can require the same Gaussian's attributes to be sent to multiple GPUs and its partial gradients to be returned from each.
Meanwhile, Both stages transfer substantial amounts of data, with backward gradient communication accounting for 45.0\% of the total volume.
Here, forward gradient communication cannot be avoided because it transfers necessary Gaussian attributes for rendering.
Thus, reducing backward gradient communication can improve the scalability of distributed 3DGS training.

\vspace{2pt}
\PBox{\textbf{Insight 4:} \textit{A substantial fraction of communicated partial gradients are zero.}}
\vspace{2pt}

We further examine the partial gradients transferred during backward gradient communication.
We classify a partial gradient as zero when all gradient fields of a Gaussian are exactly zero.
As shown in \Fig{fig:gradient}, up to 72.0\% of partial gradients are zero.
This sparsity remains consistently high across different datasets.
The high zero ratio is because an overlap between a projected Gaussian and an assigned image tile does not guarantee that this Gaussian contributes to any pixel processed by that rendering GPU.
Early termination in rasterization can prevent the Gaussian from contributing to the rendered pixels and their loss.
Thus, their partial gradient is zero.
Since accumulating zero gradients does not affect the Gaussian update, these transfers consume interconnect bandwidth without affecting training.

Together, these observations expose two major opportunities to reduce data movement in distributed 3DGS training.
First, \proj fuses forward and backward rasterization to eliminate redundant off-chip memory accesses.
Second, \proj elides zero-valued partial gradients to reduce backward inter-GPU communication.

\section{Fused Rasterization}
\label{sec:fused}

As discussed in \Sect{sec:motiv}, forward rasterization and backward rasterization account for a large fraction of the training time.
Meanwhile, existing pipelines execute them as separate GPU kernels.
Consequently, intermediate results must be repeatedly written back to off-chip memory and reloaded by subsequent kernels.
To eliminate this intra-GPU data movement, \proj proposes a fused rasterization kernel.

\subsection{Issues of the Existing Pipeline}
\label{sec:fused:issue}

\begin{figure}[t]
\centering
\subfloat[Off-chip memory traffic in the conventional 3DGS pipeline.]{
	\label{fig:org_memory_access}
    \includegraphics[width=0.48\columnwidth]{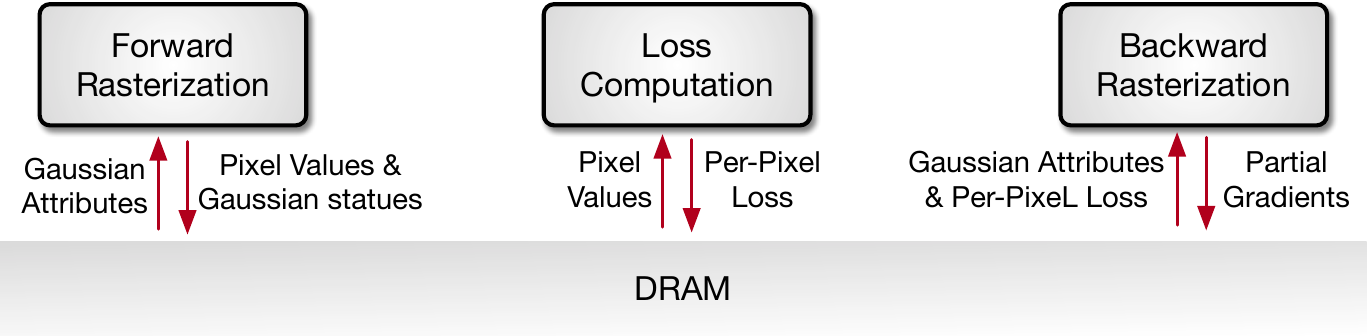}
    }
\vspace{5pt}
\subfloat[Off-chip memory traffic in \proj fused rasterization.]{
	\label{fig:our_memory_access}
	\includegraphics[width=0.48\columnwidth]{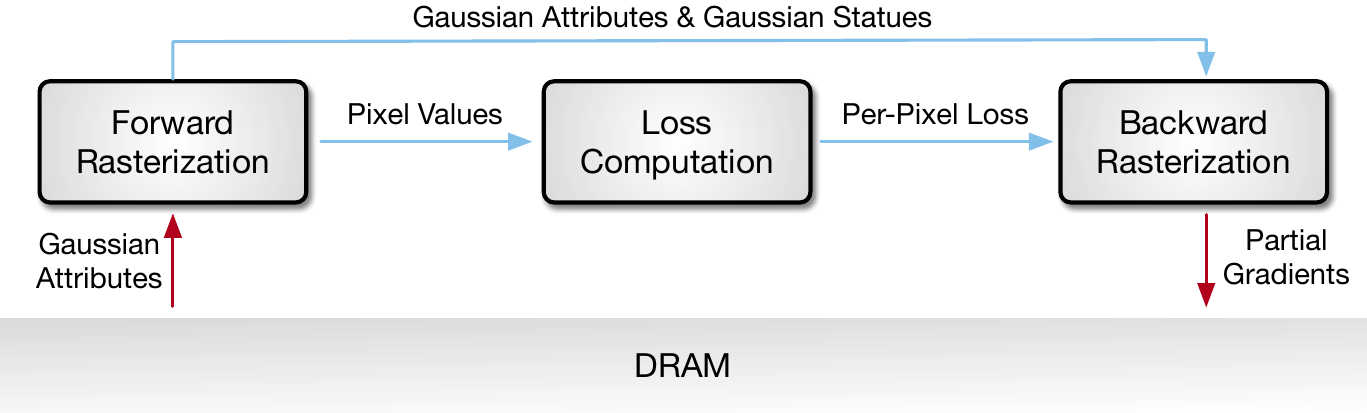}
} 
\caption{Off-chip memory traffic comparison between the conventional 3DGS pipeline and our proposed fused rasterization in \proj. Our fused rasterization eliminates the unnecessary intermediate data movements.}
\label{fig:memory_access_comp}
\end{figure}

\Fig{fig:org_memory_access} illustrates the dataflow of the original pipeline.
Forward rasterization processes each tile by loading its tile-Gaussian list and the corresponding Gaussian attributes from off-chip memory into on-chip shared memory.
Since a tile can contain more Gaussians than shared memory can accommodate, the kernel loads and processes these Gaussians in batches.
The kernel then performs front-to-back alpha blending and writes the rendered pixel colors required by the backward rasterization to off-chip memory.

Then, loss computation is launched as a separate kernel after forward rasterization.
This kernel reloads the rendered pixel colors from off-chip memory and compares them with the ground-truth image to compute the L1 and SSIM losses.
It then writes the pixel-wise loss values and per-pixel gradients, $\partial\cL/\partial\mathbf{C}(\mathbf{p})$, back to off-chip memory.

Lastly, backward rasterization is launched as a standalone kernel.
For each tile, it reloads the tile-Gaussian list, the Gaussian attributes, and the per-pixel gradients from off-chip memory.
The kernel then traverses the tile-Gaussian list in reverse order and computes the partial gradient of each pixel-Gaussian pair.
Finally, it writes the partial Gaussian gradients to off-chip memory for subsequent gradient aggregation.

As shown in \Fig{fig:org_memory_access}, this execution paradigm introduces substantial redundant memory traffic.
In particular, the Gaussian attributes are loaded twice in both forward and backward rasterization.
Moreover, intermediate results, e.g., the rendered pixels and per-pixel gradients, are written to off-chip memory by one kernel only to be immediately reloaded by another.
These redundant memory accesses make both rasterization stages memory-bound.

\subsection{Fused Rasterization Dataflow}
\label{sec:fused:method}

\para{Overview.}
To address this inefficiency, \proj fuses forward rasterization, loss computation, and backward rasterization into a single mega-kernel, as shown in \Fig{fig:our_memory_access}.
For each tile, the fused kernel loads the relevant Gaussian attributes from off-chip memory and retains them on-chip across the forward and backward rasterization.
After forward rasterization produces the rendered pixels, the kernel immediately computes the corresponding per-pixel gradients.
These intermediate gradients remain on-chip and are directly consumed by backward rasterization without being written back to off-chip memory.
Backward rasterization then computes the partial Gaussian gradients and writes only the outputs to off-chip memory for gradient aggregation.

This fused dataflow eliminates the round trips of rendered pixels, intermediate rasterization states, and per-pixel gradients through off-chip memory.
It also allows the forward and backward rasterization to reuse Gaussian attributes while avoiding redundant attribute loading.

\para{Tile-Local Loss Computation.}
One caveat here is that fusing loss computation with rasterization is challenging.
Because the conventional SSIM loss introduces dependencies across image tiles.
SSIM loss computes the structural similarity over a spatial window surrounding each pixel.
For pixels near a tile boundary, this window can contain pixels processed by other thread blocks or rendering GPUs.
A thread block therefore cannot compute SSIM loss using only the pixel values generated for its own tile.

To eliminate these dependencies, \proj adopts the stochastic structural similarity loss, S3IM~\cite{xie2023s3im}, and applies it independently within each image tile.
For a tile $t$, \proj computes the training loss as,
\begin{equation}
    \cL_t
    =
    (1-\lambda)
    \cL_{1}\!\left(\mathbf{C}_t,\mathbf{C}^{*}_t\right)
    +
    \lambda
    \cL_{\mathrm{S3IM}}\!\left(\mathbf{C}_t,\mathbf{C}^{*}_t\right),
    \label{eq:tile-local-loss}
\end{equation}
where $\mathbf{C}_t$ and $\mathbf{C}^{*}_t$ denote the rendered and ground-truth pixel colors of tile $t$, respectively.
The $\cL_{\mathrm{S3IM}}$ term makes its structural comparisons using only pixels belonging to the same tile.
Consequently, each thread block can compute its pixel-wise gradients using only its locally generated pixels.

After forward rasterization, the fused kernel retains the rendered pixels in shared memory and directly evaluates the loss function.
It then keeps the pixel-wise gradients on-chip for immediate consumption by backward rasterization.
Because the computation has no dependencies on neighboring tiles, it requires neither cross-block synchronization nor boundary-data exchange between GPUs.
Our evaluation in \Sect{sec:eval} shows that the tile-local loss preserves the same reconstruction quality as the original implementation.
\section{Hardware-Aware Training}
\label{sec:training}

Although fused rasterization eliminates substantial off-chip memory traffic, it introduces a new challenge: increased on-chip resource pressure.
In this section, we first explain why applying fused rasterization to all image tiles is impractical.
We then present a novel hardware-aware training that increases the fraction of tiles that can benefit from fusion.

\subsection{Challenges}
\label{sec:training:ch}

In the conventional pipeline, forward rasterization, loss computation, and backward rasterization execute as separate GPU kernels, as shown in \Fig{fig:org_memory_access}.
Between kernels, intermediate results have to be written to off-chip memory and subsequently reloaded by the consumer kernel.
Thus, this execution model incurs substantial off-chip memory traffic.

Fused rasterization makes the opposite tradeoff.
It retains all intermediate data, e.g., Gaussian attributes and pixel values, on-chip until backward rasterization completes, e.g., \Fig{fig:our_memory_access}.
Keeping the data on-chip eliminates intermediate off-chip accesses but substantially increases the shared-memory pressure of the fused kernel.
Moreover, the shared-memory requirement varies across thread blocks.
Because each thread block processes one image tile, its on-chip memory requirement depends on how many Gaussians intersect with that tile.
A tile intersected by more Gaussians requires the fused kernel to retain a larger amount of data on-chip.

If the shared-memory requirement of a tile exceeds the per-block capacity, that tile cannot execute through the fused kernel.
Even when the requirement remains below the hardware limit, a large per-block shared-memory allocation can reduce the number of concurrently resident thread blocks on each streaming multiprocessor.
This could weaken the GPU's ability to overlap compute and memory latency, potentially offsetting its benefits of reduced off-chip traffic.

Therefore, directly applying fused rasterization to every tile is neither practical nor efficient.
\proj must determine which tiles can execute efficiently within the shared-memory budget of the target GPU.
Tiles that exceed this budget must follow the conventional execution path and store their intermediate data back to off-chip memory.
Thus, we introduce a training strategy in \Sect{sec:training:method} to maximize fusion coverage while preserving reconstruction quality.

\begin{figure*}[t]
    \centering
    \includegraphics[width=\textwidth]{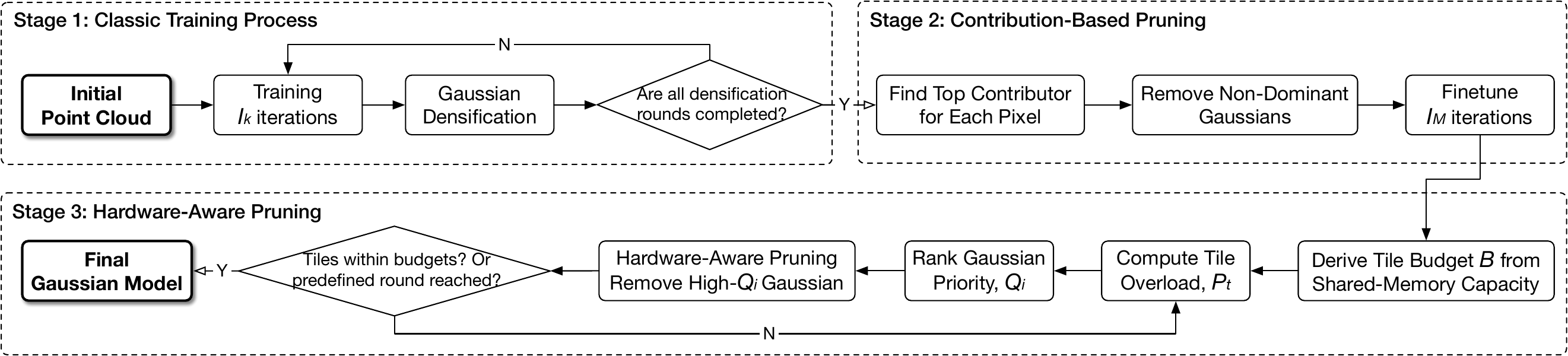}
    \caption{Overview of \proj's three-stage training strategy.
    Stage~1 follows the classic 3DGS training process.
    Stage~2 then identifies the dominant Gaussian contributor for each training pixel, removes Gaussians that never become a dominant contributor, and fine-tunes the remaining representation.
    Stage~3 computes the per-tile Gaussian budget $B$ from the shared-memory capacity of the target GPU, calculates the tile overload $P_t$, and ranks Gaussians using the pruning priority $Q_i$.
    It then iteratively removes high-priority Gaussians until all tiles satisfy their budgets or the predefined pruning-round limit is reached.}
\label{fig:hardware-aware-training}
    \label{fig:strategy}
\end{figure*}

\subsection{Training Strategy}
\label{sec:training:method}

To alleviate shared-memory pressure, \proj proposes a three-stage training strategy in \Fig{fig:strategy} that increases the fraction of tiles eligible for fused rasterization.
The first stage follows the classic 3DGS training procedure to construct an initial Gaussian representation.
Then, the second stage removes insignificant Gaussians that make no major contribution to any image in the training set.
Lastly, the final stage accounts for the GPU shared-memory capacity and progressively removes unimportant Gaussians from overloaded tiles.
Thus, more tiles can leverage fused rasterization.

\para{Stage 1: Initial Training and Densification.}
Starting from the initial point cloud, \proj first trains a Gaussian representation using the classic 3DGS training procedure.
At each predefined iteration interval, $I_k$, it performs densification to add Gaussians in scene regions that require more fine-grained representations.
This training process is repeated for multiple rounds until the overall training reaches the predefined training iteration, $I_N$.
At this stage, \proj does not impose any hardware-dependent constraints.

\para{Stage 2: Contribution-Based Initial Pruning.}
After standard training, \proj first removes Gaussians that make no major contribution to any rendered pixel.
Specifically, we identify the Gaussian with the largest rendering contribution to each pixel $\mathbf{p}$ in the training set,
\begin{equation}
    i^{*}(\mathbf{p})
    =
    \arg\max_{i\in\cI(\mathbf{p})} (T_i \alpha_i \mathbf{c}_i),
    \label{eq:dominant_gaussian}
\end{equation}
where $\cI(\mathbf{p})$ is the set of intersected Gaussians contributing to pixel $\mathbf{p}$, and $(T_i \alpha_i \mathbf{c}_i)$ is the rendering contribution of Gaussian $G_i$ to that pixel.
$i^{*}(\mathbf{p})$ is the index of that important Gaussian.
We then retain the union of the important Gaussians, $\cG_{\mathrm{i*}}$, selected by all pixels across the training set.

This initial pruning stage is independent of the target hardware.
Its purpose is to eliminate redundant Gaussians before applying the more aggressive hardware-aware pruning.
After initial pruning, \proj further finetunes the remaining Gaussians for $I_M$ iterations without densification so that they can compensate for the removed Gaussians.

\para{Stage 3: Hardware-Aware Pruning.}
After finetuning, \proj further prunes the Gaussian representation based on the shared-memory capacity of the target GPU.
We first translate the shared-memory constraint into the maximum number of Gaussians that can be processed by one fused thread block.
Here, $M_{\mathrm{budget}}$ is the shared-memory budget assigned to each thread block, $M_{\mathrm{fixed}}$ is the fixed storage requirement per tile, and $M_{\mathrm{g}}$ is the on-chip storage required for each Gaussian.
The Gaussian budget of each tile is,
\begin{equation}
    B
    =
    \left\lfloor
        \frac{
            M_{\mathrm{budget}}-M_{\mathrm{fixed}}
        }{
            M_{\mathrm{g}}
        }
    \right\rfloor.
    \label{eq:tile_budget}
\end{equation}
Because $M_{\mathrm{budget}}$ and $M_{\mathrm{fixed}}$ can be determined offline by the target GPU and the implementation, the budget $B$ explicitly reflects the underlying hardware configuration.

For each tile $t$, \proj then counts the number of intersected Gaussians, denoted by $N_t$.
The number of excess Gaussians in that tile can be expressed as,
\begin{equation}
E_t = \max\left( 0, N_t-B \right).
\label{eq:excess_gaussians}
\end{equation}
A tile is eligible for fused rasterization when $E_t=0$, and it is overloaded when $E_t>0$.

To reduce the number of overloaded tiles, we first propose a metric to quantify the severity of this overload using the tile-level overload pressure, $P_t$:
\begin{equation}
P_t = \frac{\max\left(0,N_t-B\right)}{N_t}.
\label{eq:tile_pressure}
\end{equation}
$P_t$ is $0$ when $N_t = 0$.
A larger $P_t$ means that a larger fraction of the Gaussians intersecting tile $t$ must be removed before this tile can satisfy the hardware budget.

Meanwhile, one Gaussian can intersect multiple tiles.
We need to determine which Gaussian should be removed to achieve the maximal shared-memory pressure reduction across multiple thread blocks simultaneously.
We therefore define the overload pressure of Gaussian $G_i$ as,
\begin{equation}
P_i = \sum_{t\in\cT_i} P_t,
\label{eq:gaussian_pressure}
\end{equation}
where $\cT_i$ is the set of tiles intersected by Gaussian $G_i$.
We assign a high pressure score, $P_i$, to a Gaussian that contributes to many severely overloaded tiles.

However, hardware pressure alone cannot determine which Gaussians to remove because we must preserve visually important Gaussians.
Thus, we define $I_i$ to be the normalized relative importance of Gaussian $G_i$, derived from its rendering contribution.
We calculate its pruning priority as,
\begin{equation}
Q_i = \frac{P_i}{I_i+\epsilon},
\label{eq:hardware-aware-pruning-priority}
\end{equation}
where $\epsilon$ is a small constant that prevents numerical instability.
A Gaussian receives a high pruning priority when it has high contributions to the overloaded tiles but has low relative importance to reconstruction quality.

During our last stage, \proj performs several rounds of pruning.
Each time, \proj ranks the candidate Gaussians in descending order of $Q_i$.
It then removes the $\beta$ percent highest-priority candidates from the overall Gaussian representation.
Next, \proj would finetune $I_o$ iterations.  
The pruning process continues until the targeted tiles satisfy their Gaussian budgets or the predefined round of pruning is reached.
This way, we can remove low-importance Gaussians whose removal creates the greatest number of additional fusion opportunities.

Unlike conventional pruning methods that primarily minimize the global model size, our strategy directly reshapes the per-tile Gaussian distribution according to the shared-memory capacity of the target GPU.
The results in \Sect{sec:eval:abl} show that our training process increases the fraction of fusion-eligible tiles from 33.2\% to 50.7\% while preserving reconstruction quality, compared to prior pruning methods.

\section{Zero-Eliding Gradient Communication}
\label{sec:comm}

As discussed in \Sect{sec:motiv}, the gradient aggregation in the backward pass dominates the inter-GPU communication latency of distributed 3DGS training.
Existing systems communicate all partial gradients of every Gaussian processed by a rendering GPU, regardless of the value of that gradient.
However, our characterization shows that up to 72.0\% of these gradients are zeros.
This motivates \proj to eliminate zero-value gradients before communication.

We first present \textit{Zero-Gradient Elision} that identifies and removes zero-value gradients without changing the training semantics.
Then, we present an efficient \textit{Sparse Gradient Transmission} for effective encoding and communication.

\subsection{Zero-Gradient Elision}
\label{sec:comm:elision}

As described in \Sect{sec:bg:pipeline}, projection associates a Gaussian with every intersecting image tile.
However, a tile-Gaussian intersection does not guarantee that this Gaussian contributes to any pixel in this image tile.
This is because a Gaussian may be skipped by a pixel when the accumulated transmittance of the pixel reaches the early-termination threshold.
Consequently, many Gaussians communicated to a rendering GPU do not contribute to any pixel assigned to that GPU.
Naturally, those skipped Gaussians will obtain zero gradient during the gradient backpropagation.

However, existing distributed 3DGS systems use the same indexing during forward Gaussian communication to return partial gradients to owner GPUs.
This symmetric indexing simplifies gradient routing because the owner GPU can recover the destination Gaussian directly from the original communication order.
However, it also forces the rendering GPU to return gradients of every Gaussian, including those whose partial gradients are zero.

To avoid this redundant data communication, \proj examines each partial-gradient record after backward rasterization and flags those Gaussians whose gradients are zero.
After flagging, \proj removes these zero-value gradients from the reverse-communication buffer without deleting the corresponding Gaussians from the model.
It retains only the Gaussians and their partial gradients containing at least one nonzero gradient value.
Because every omitted partial gradient is exactly zero, this optimization preserves the result of gradient aggregation and does not affect training semantics.
By avoiding these redundant gradients, \proj reduces both the inter-GPU communication volume and the local gradient aggregation overhead.
The next subsection describes how \proj efficiently packs and transmits the remaining nonzero gradient records.

\begin{figure}[t]
    \centering
    \includegraphics[width=\columnwidth]{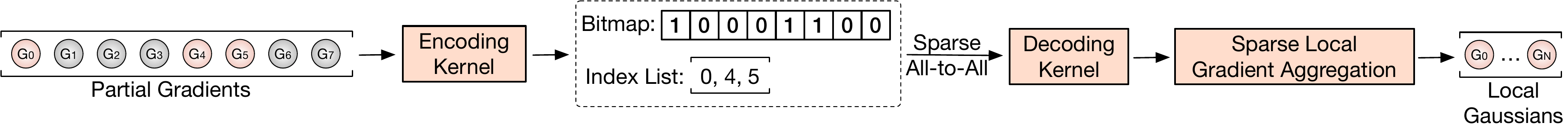}
    \caption{Overview of zero-gradient elision and sparse gradient transmission in \proj. The rendering GPU packs nonzero gradients using a bitmap, and the owner GPU decodes and directly aggregates the sparse gradients.}
    \label{fig:sparse_comm}
\end{figure}

\subsection{Sparse Gradient Transmission}
\label{sec:comm:transmission}

To support zero-gradient elision, \proj introduces a GPU-efficient encoding and sparse communication scheme.

\para{Encoding Scheme.}
In \Fig{fig:sparse_comm}, after backward rasterization, \proj first groups the partial gradients according to their destination owner GPUs, and then divides each partition into fixed-size chunks.
The encoder assigns one thread block to each chunk and one thread to a subset of gradients in a chunk.
Each thread checks whether the gradients are nonzero, and the threads collectively construct a bitmap that identifies the nonzero gradients.
The thread block then performs an on-chip prefix sum to determine the packed position of each nonzero gradient.
Finally, the encoder writes the nonzero gradients contiguously into a compact array.

\para{Sparse Communication and Local Aggregation.}
After encoding, the rendering GPUs exchange their compact array with the corresponding bitmap and perform a variable-length sparse all-to-all communication.
On each owner GPU, the decoder assigns each received chunk to one thread block and uses its bitmap to recover the original positions of the retained gradients.
The decoder then uses the forward communication indexing to map those partial gradients to its local Gaussian indices.
Rather than reconstructing the original dense gradient array, the decoder directly accumulates each nonzero gradient into the corresponding Gaussian gradient buffer.
Because a Gaussian may receive gradients from multiple rendering GPUs, these updates use atomic operations to ensure correctness.
By completely skipping the zero-value gradients, this design reduces the workloads in both inter-GPU communication and local aggregation.

\section{Experimental Setup}
\label{sec:exp}

\para{Hardware platforms.}
To evaluate the robustness of our training framework, we evaluate \proj on two GPU servers with different interconnect configurations.
One is an A6000 server with eight Nvidia A6000 GPUs. 
Each has 48~GB GDDR6 memory.
The other one is an A100 server with eight Nvidia A100 GPUs.
Each has 80~GB HBM memory.
We then measure the end-to-end performance of distributed training, including the kernel launch time.

\para{Datasets.}
Our evaluation covers six widely-used datasets, including Mill-19~\cite{turki2022mega}, UrbanScene3D~\cite{lin2022capturing}, MatrixCity~\cite{li2023matrixcity}, Mip-NeRF 360~\cite{barron2022mipnerf360}, DeepBlending~\cite{hedman2018deep}, Tanks\&Temples~\cite{Knapitsch2017}.
The first three datasets are for large-scale scene reconstruction, including various urban scenes.
The remaining scenes are well-known 3DGS datasets with different indoor and outdoor scenes.
For both training and evaluation, we all use the native image resolution.
For testing, we reserve every eighth view in the dataset, following the standard 3DGS protocol~\cite{zhao2025on}.

\para{Training Settings.}
We train every scene at its native resolution on eight GPUs with one camera
per GPU. Because the scenes differ substantially in scale, we do not terminate initial
training at a common iteration count.  Instead, we resume densification from
intermediate checkpoints until the Gaussian population and validation quality
stabilize, and then disable densification and opacity reset for a final
convergence phase.  Across our workloads, initial training takes approximately
200k iterations on average, with the exact stopping point adjusted to scene
scale and convergence.  We apply pruning only after this initial model has converged.  The retained Gaussians
are then fine-tuned for another 30k iterations with densification disabled;
all quality and performance results use the checkpoint at the end of this
fine-tuning phase.  The unpruned and pruned runs otherwise use the same data
split, image resolution, loss, and optimizer hyperparameters.

\para{Metrics.}
We evaluate training performance using end-to-end latency and speedup, and reconstruction quality using PSNR and SSIM.
To assess our \textit{hardware-aware training}, we report the pruning ratio, defined as the fraction of Gaussians removed from the original training.
Meanwhile, we also report the percentage of image tiles whose Gaussian working sets fit within the fused-kernel resource budget before and after applying our pruning technique.

\para{Variants.}
Our evaluation also includes several variants to dissect the contributions in \proj:
\begin{itemize}
    \item \mode{Base}: this variant follows the techniques in the original Grendel-GS~\cite{zhao2025on}.
    \item \mode{Base+Fused}: this variant only applies our fused rasterization on top of \mode{Base}.
    \item \mode{Base+Fused+SpComm}: this variant applies both our fused rasterization and our sparse communication.
    \item \mode{\proj}: this one is our full-fledged implementation of \proj design with all proposed techniques.
\end{itemize}
\section{Evaluation}
\label{sec:eval}

\subsection{Rendering Quality}
\label{sec:eval:acc}

\begin{figure*}[t]
    \centering
    \subfloat[PSNR. Higher is better.\label{fig:quality:psnr}]{%
        \includegraphics[width=0.48\textwidth]{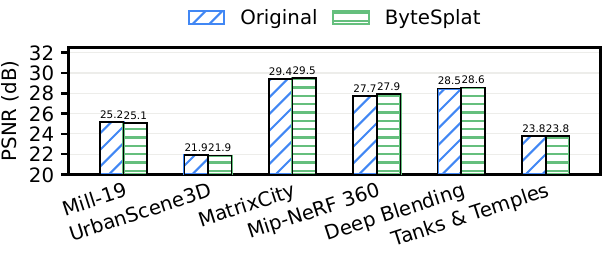}%
    }
    \hfill
    \subfloat[SSIM. Higher is better.\label{fig:quality:ssim}]{%
        \includegraphics[width=0.48\textwidth]{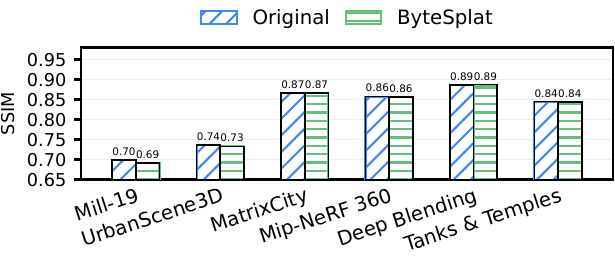}%
    }
    \caption{Rendering quality before and after hardware-aware pruning.}
    \label{fig:quality}
\end{figure*}

\Fig{fig:quality} shows the overall quality comparison between the original Grendel-GS framework~\cite{zhao2025on} and \proj with hardware-aware training.
Specifically, \Fig{fig:quality:psnr} shows the PSNR results of two methods over six different datasets.
Overall, \proj achieves similar visual quality with virtually no PSNR degradation.
The largest degradation is only 0.1~dB, while the average PSNR increases by 0.04~dB.
Notably, several scenes improve the quality slightly after pruning.
This shows that our hardware-aware training allows the remaining Gaussians to accommodate the contribution of removed Gaussian primitives.

Similarly, \Fig{fig:quality:ssim} shows the same trend in SSIM.
On average, Grendel-GS achieves 0.82 in SSIM, while \proj achieves 0.81.
The largest SSIM decrease is 0.01.
Together, the PSNR and SSIM results show that \proj can reshape the Gaussian distribution for fused execution without causing a noticeable loss in rendering quality.

\subsection{Performance}
\label{sec:eval:gpu}

\begin{figure*}[t]
    \centering
    \subfloat[Eight Nvidia A6000 GPUs.\label{fig:perf_a6000}]{%
        \includegraphics[width=0.441\textwidth]{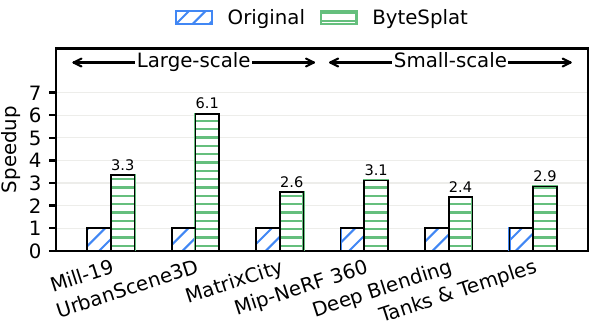}%
    }
    \hfill
    \subfloat[Eight Nvidia A100 GPUs.\label{fig:perf_h100}]{%
        \includegraphics[width=0.441\textwidth]{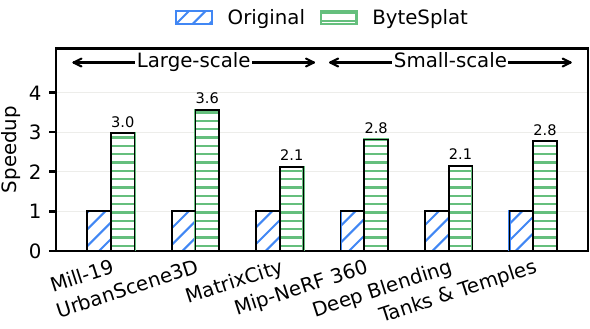}%
    }
    \caption{End-to-end per-iteration training speedup. We normalize all numbers to the corresponding Grendel-GS baselines.}
    \label{fig:perf}
\end{figure*}

\begin{figure*}[t]
    \begin{minipage}[t]{0.32\textwidth}
        \centering
        \includegraphics[height=0.13\textheight]{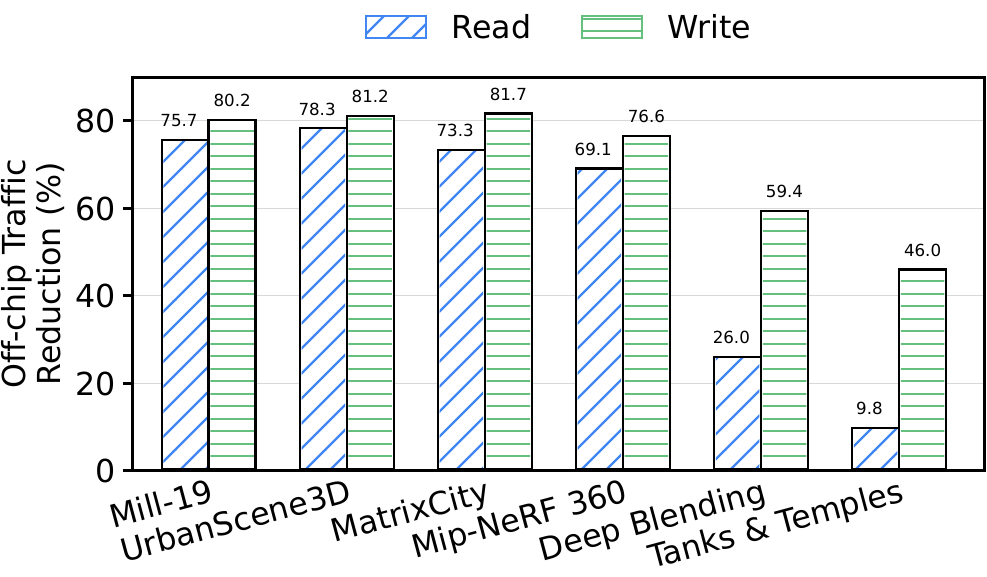}
        \captionof{figure}{Reduction of Off-chip Memory Accesses.}
        \label{fig:memory-access-reduction}
    \end{minipage}\hfill
    \begin{minipage}[t]{0.32\textwidth}
        \centering
        \includegraphics[height=0.13\textheight]{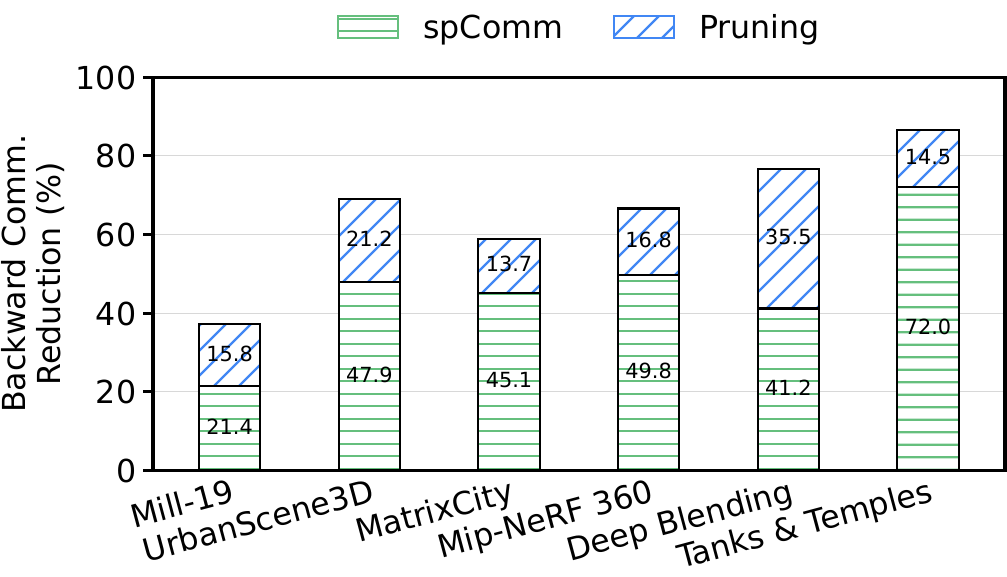}
        \captionof{figure}{Reduction of Inter-GPU Communication.}
        \label{fig:backward-communication-reduction}
    \end{minipage}\hfill
    \begin{minipage}[t]{0.32\textwidth}
        \centering
        \includegraphics[height=0.13\textheight]{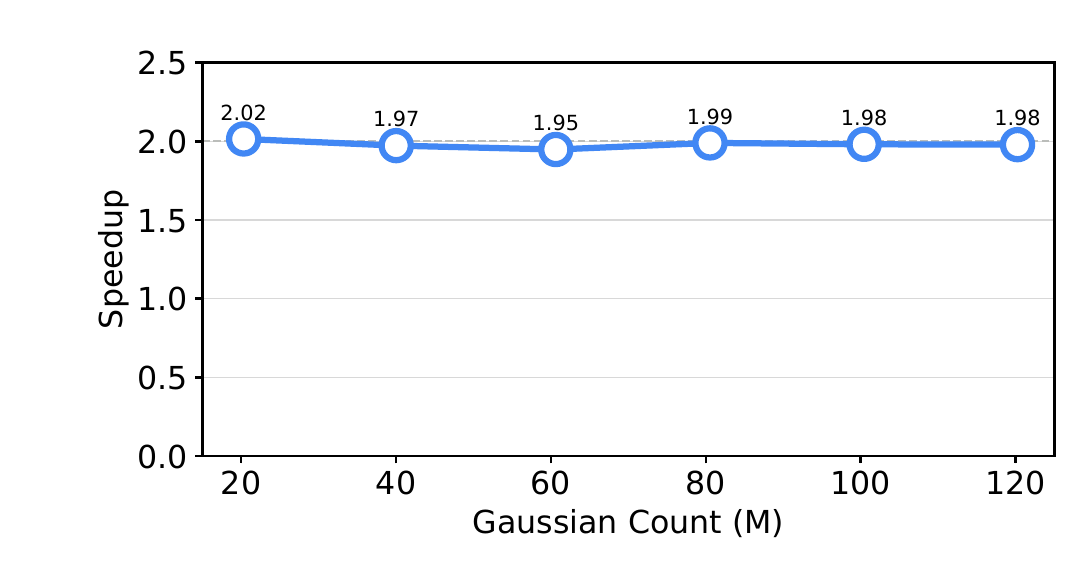}
        \captionof{figure}{Training Speedup with Increasing Gaussian Count.}
        \label{fig:gaussian-count-speedup}
    \end{minipage}
\end{figure*}

\para{Overall Performance.}
Next, we show the performance comparison between Grendel-GS and \proj in \Fig{fig:perf}.
To show the robustness of \proj, we measure the performance on two different server platforms: an A6000 server with eight Nvidia A6000 GPUs in \Fig{fig:perf_a6000} and an A100 server with eight Nvidia A100 GPUs in \Fig{fig:perf_h100}.
Here, we measure the per-iteration latency and calculate the relative speedup of \proj compared to Grendel-GS.
On the A6000 server, \proj achieves an average speedup of 3.4$\times$ and up to 6.1$\times$ over the original pipeline.
On A100, the corresponding average and maximum speedups are 2.7$\times$ and 3.6$\times$.
These improvements include all three sources of data-movement reduction: fused rasterization avoids redundant DRAM traffic, hardware-aware pruning increases the fraction of tiles that use the fused path, and zero-gradient elision reduces backward communication and aggregation.
\Sect{sec:eval:abl} further shows an ablation study of \proj on an A6000 server to show the contribution of each component.

\para{Cross-Platform Performance.}
\proj improves training performance across both large-scale and small-scale scenes.
The variation across scenes shows two things: 1) the amount of work removed by pruning and 2) the fraction of the original iteration spent in rasterization and gradient aggregation.
Dense scenes often provide more opportunities to reduce data movement, while pruned bounded scenes can also obtain large gains.
The consistent performance gains on A6000 and A100 servers further show that \proj is not tied to one particular GPU compute resource, memory bandwidth, or interconnect performance.

\para{Performance with Various Model Sizes.}
We also evaluate whether the performance benefit of \proj persists as the Gaussian representation grows beyond the scale of our main evaluation.
Specifically, we keep the training configuration unchanged and increase the representation from 20.3 million to 120.2 million Gaussians on the A100 server.
As shown in \Fig{fig:gaussian-count-speedup}, \proj maintains a stable 1.9-2.0$\times$ speedup over the Grendel-GS implementation across all six model sizes under the same Gaussian counts, even as the representation grows over 100 million Gaussians.
At the largest scale, \proj reduces the per-iteration latency from 114.4 to 57.8~ms.
The consistent speedup confirms that the reduction in redundant data movement continues to benefit training as the Gaussian workload increases.

\para{Reduction of Off-chip Memory Accesses.}
To further show the source of our performance gain, we first show the reduction of off-chip memory accesses.
This shows why our fused rasterization can improve performance.
Specifically, we use GPU performance counters to compare \mode{Base} with \mode{Base+Fused} while keeping the Gaussian representation and all other training settings unchanged.
We report off-chip reads and writes separately in \Fig{fig:memory-access-reduction} because fused rasterization affects them in different ways.
Across the six datasets, it reduces read traffic by 55.4\% on average and by up to 78.3\%, mainly by avoiding repeated loads of Gaussian attributes and intermediate states.
It also reduces write traffic by 70.9\% on average and by up to 81.7\%, as rendered colors and loss gradients no longer need to be stored between separate stages.
The consistent reductions in both directions confirm that retaining reusable data on chip removes a large fraction of the data movement in the original pipeline.

\para{Reduction of Inter-GPU Communication.}
In addition, we further show the effectiveness of both hardware-aware training and zero-gradient elision by showing the inter-GPU communication reduction.
Specifically, \Fig{fig:backward-communication-reduction} shows the inter-communication reductions in the backward pass.
Each stacked bar normalizes the reduction to the communication volume of the original pipeline in Grendel-GS.
First, sparse communication removes zero gradients from the unpruned representation, reducing traffic by 46.2\% on average and by up to 72.0\% across the six datasets.
Applying hardware-aware pruning next contributes a further 19.6\% on average and up to 35.5\% by reducing the unnecessary Gaussian points.
Together, the two stages reduce backward communication by 65.8\% on average and by up to 86.5\% relative to the original pipeline.
The benefits are complementary: hardware-aware pruning reduces the number of Gaussians in communication, while zero-gradient elision removes the remaining records that do not affect the model update.
Because \proj elides only exactly zero gradients, this communication optimization does not change the final rendering quality.

\subsection{Ablation Study}
\label{sec:eval:abl}

\begin{figure*}[t]
    \centering
    \includegraphics[width=0.82\textwidth]{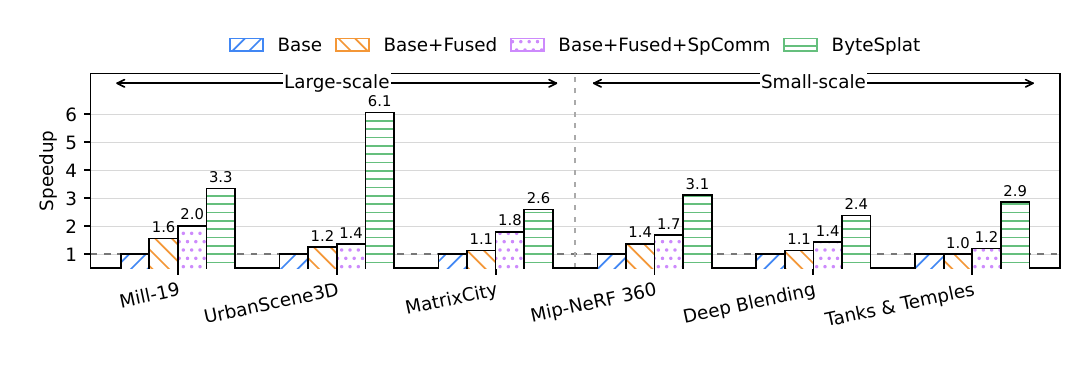}
    \caption{The ablation study of per-iteration performance on the four variants shown in \Sect{sec:exp}. All performance numbers are normalized to the \mode{Base} variant. The performance is measured on an A6000 server with eight A6000 GPUs.}
    \label{fig:cumulative-ablation}
\end{figure*}

\para{Setup.}
We use a cumulative ablation study to show how each major component contributes to the end-to-end performance of \proj.
All configurations use the same eight-GPU A6000 server, camera order, and training setup as the end-to-end experiment.
\mode{Base} follows the original Grendel-GS pipeline;
\mode{Base+Fused} adds fused rasterization;
\mode{Base+Fused+SpComm} further enables sparse gradient communication while retaining the same unpruned models; and \mode{\proj} evaluates the full-fledged system on the corresponding hardware-aware pruned models.
Every bar shows speedup relative to \mode{Base} on the original unpruned Gaussian model.

\para{Overall Performance.}
As shown in \Fig{fig:cumulative-ablation}, across the six datasets, \mode{Base+Fused} achieves a geometric-mean speedup of 1.2$\times$.
By introducing zero-eliding elision, \mode{Base+Fused+SpComm} boosts the speedup to 1.6$\times$, i.e., a further 1.3$\times$ improvement over \mode{Base+Fused}.
Applying the full-fledged system, \mode{\proj} raises the end-to-end speedup to 3.2$\times$, which is 2.1$\times$ faster than \mode{Base+Fused+SpComm}.

The benefits also vary with scene scale.
For the three large-scale datasets, the complete \mode{\proj} runtime achieves 3.7$\times$ geometric-mean speedup, compared with 2.8$\times$ for the three small-scale datasets.
Fused rasterization, sparse communication, and hardware-aware pruning provide speedups of 1.3$\times$, 1.3$\times$, and 2.2$\times$ on large-scale datasets, respectively.
The corresponding increments on small-scale datasets are 1.2$\times$, 1.2$\times$, and 1.9$\times$.
Overall, these results demonstrate that \proj is particularly effective for
large-scale scenes, whose larger Gaussian populations expose more opportunities
for eliminating redundant computation and data movement.

\begin{figure*}[t]
    \centering
    \begin{minipage}[t]{0.48\textwidth}
        \centering
        \includegraphics[width=\textwidth]{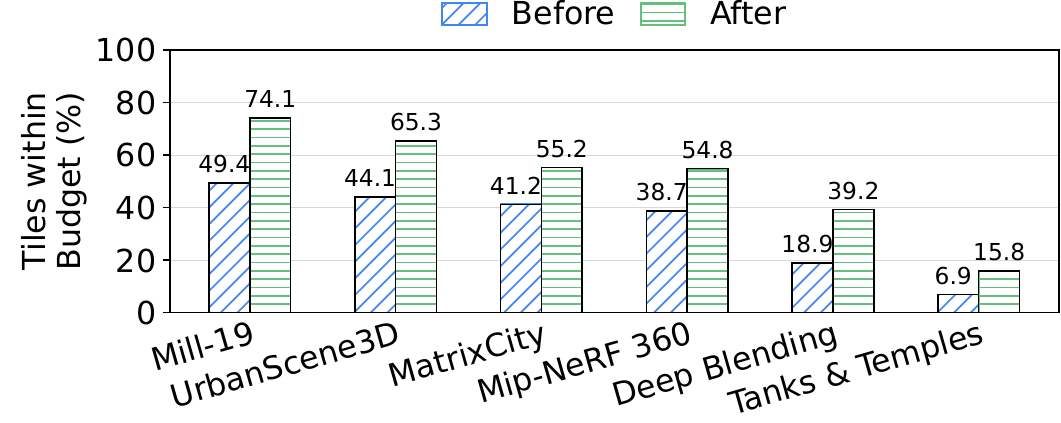}
        \caption{Fraction of tiles satisfying the Gaussian budget before and after applying our hardware-aware training.}
        \label{fig:tile-budget-satisfaction}
    \end{minipage}
    \hfill
    \begin{minipage}[t]{0.48\textwidth}
        \centering
        \includegraphics[width=\textwidth]{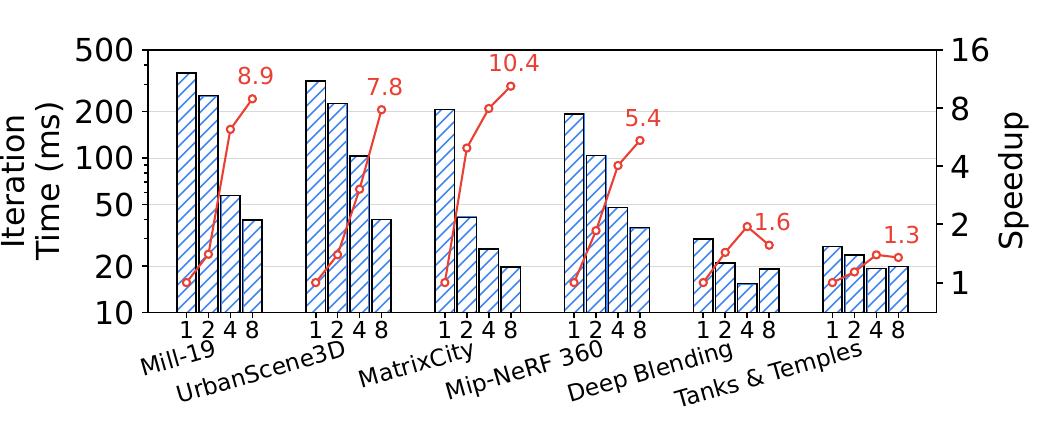}
        \caption{The scalability analysis. The per-iteration latency of \proj as the GPU count increases.}
        \label{fig:optimized-training-scaling}
    \end{minipage}
\end{figure*}

\para{Tile-Budget Satisfaction.}
\Fig{fig:tile-budget-satisfaction} further compares the fraction of tiles that satisfy the fused-kernel Gaussian budget before and after hardware-aware pruning.
Pruning improves this fraction on all six datasets, with gains ranging from 8.9\% to 24.7\% and an average gain of 17.5\%.
The improvement holds for both large- and small-scale scenes, showing that the training strategy consistently reshapes crowded tiles rather than merely reducing the total number of Gaussians.
As a result, more tiles can use fused rasterization and benefit from its lower off-chip traffic.

\subsection{Scalability Analysis}
\label{sec:eval:scale}

\para{Multi-GPU Performance.}
\Fig{fig:optimized-training-scaling} shows the per-iteration latency of \proj as the GPU count increases from one to eight.
Using eight GPUs provides an average speedup of 5.9$\times$ and a maximum speedup of 10.4$\times$ over a single GPU.
The three large-scale datasets achieve 7.8-10.4$\times$ speedup, while the small-scale scenes achieve 1.3-5.4$\times$.
The super-linear scalability is because distributed training largely reduces the frequent off-chip traffic.
Large scenes benefit more because they provide sufficient work for additional GPUs, whereas inter-communication overhead accounts for a larger fraction of the iteration time in smaller scenes.
These results show that \proj scales most effectively on the large workloads for which distributed training is essential.

\subsection{Sensitivity Study}
\label{sec:eval:sens}

\begin{figure*}[t]
    \centering
    \subfloat[Per-iteration training speedup.\label{fig:pruning-sensitivity:speedup}]{%
        \includegraphics[width=0.44\textwidth]{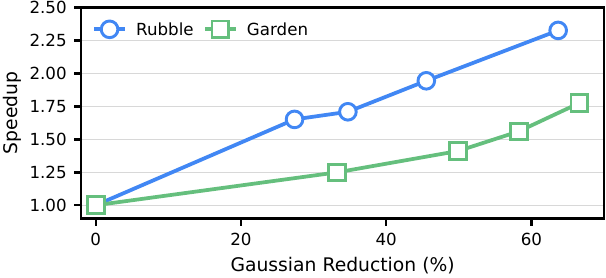}%
    }
    \hfill
    \subfloat[PSNR evaluation.\label{fig:pruning-sensitivity:psnr}]{%
        \includegraphics[width=0.44\textwidth]{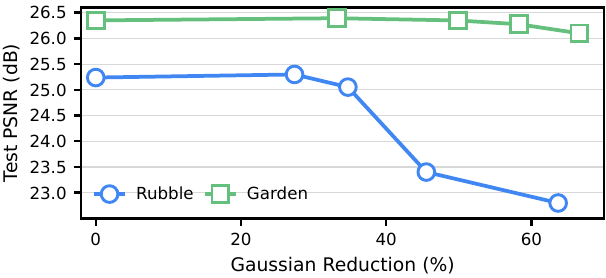}%
    }
    \caption{Sensitivity of performance and reconstruction quality to the pruning ratio in hardware-aware training. Here, we pick one large-scale dataset, Mill-19, and one small-scale dataset, Mip-NeRF 360.}
    \label{fig:pruning-sensitivity}
\end{figure*}

\para{Pruning Ratio.}
Here, we show the sensitivity of performance and reconstruction quality to the pruning ratio in hardware-aware training.
In this study, we choose two datasets, one is a large-scale dataset, Mill-19, the other is a small-scale dataset, Mip-NeRF 360.
For each scene, we start from the same converged model, then vary the fraction of pruned Gaussians, and fine-tune every pruned model for the same number of iterations.
We report both PSNR for rendering quality and end-to-end iteration latency on eight A6000 GPUs for training efficiency in \Fig{fig:pruning-sensitivity}.

On the Mill-19 dataset, increasing the pruning ratio from 0\% to 63.7\% reduces per-iteration latency from 69.2~ms to 29.8~ms, yielding a 2.3$\times$ speedup. 
However, the reconstruction quality decreases from 25.2~dB to 22.8~dB.
With light pruning at 34.7\%, the Mill-19 dataset already achieves 1.8$\times$ speedup while retaining the same reconstruction quality.
Further increasing the ratio to 45.5\% improves the speedup to 1.9$\times$, but introduces noticeable quality loss to 1.8\,dB, revealing a clear quality-efficiency turning point.

The Mip-NeRF 360 dataset reveals a smoother tradeoff.
Increasing its pruning ratio from 0\% to 66.6\% reduces iteration latency from 64.0~ms to 36.1~ms and provides a 1.8$\times$ speedup, while test PSNR decreases by only 0.3\,dB.
When removing 58.3\% of the total Gaussians, \proj reduces latency to 41.0~ms and achieves a 1.6$\times$ speedup with a 0.1\,dB quality loss.
The light and moderate settings also retain essentially unchanged reconstruction quality.
This shows that the small-scale scenes are less sensitive to the pruning ratio than large-scale scenes.

Overall, the Mill-19 dataset shows a noticeable turning point after a given pruning ratio, whereas small-scale scenes degrade more gradually.
This is because a large-scale scene often requires more Gaussians to represent its geometry and appearance.
Once pruning removes this necessary representation capacity, its reconstruction quality decreases rapidly.
However, the pruning ratios used by \proj capture most of the available efficiency improvement without entering the range where additional pruning causes significant quality loss.
\section{Related Work}
\label{sec:related}

\para{3DGS GPU Optimization.}
Existing studies improve GPU-based 3DGS efficiency through various techniques, including compact representations, finer-grained culling, and optimized rasterization.
For instance, CompactGS~\cite{lee2024compact}, EAGLES~\cite{girish2025eagles}, and Compressed 3DGS~\cite{niedermayr2024compressed} reduce representation size through compact parameter encodings and quantization.
On the other hand, HAC~\cite{chen2024hac} and HAC++~\cite{chen2025hac++} further improve compression through context-based coding.
These approaches primarily reduce model storage requirements, while another line of work optimizes runtime execution.
For example, AdR-Gaussian~\cite{wang2024adr} and Speedy-Splat~\cite{hanson2025speedy} tighten the regions processed for each Gaussian to avoid unnecessary rasterization.
FlashGS~\cite{feng2025flashgs} combines precise intersection tests, workload-aware scheduling, and software pipelining.
Balanced 3DGS~\cite{gui2024balanced} improves workload distribution through Gaussian-wise parallelism and fine-grained tiling, while TC-GS~\cite{liao2025tc} maps Gaussian evaluation to matrix operations on Tensor Cores.
Seele~\cite{zhu2026seele} combines scene preprocessing, online filtering, and contribution-aware rasterization to improve mobile GPU efficiency.

In contrast to representation compression and optimizations within rendering stages, \proj reduces repeated off-chip accesses across forward rasterization, loss computation, and backward rasterization through fused execution.

\para{3DGS Training Acceleration.}
Numerous approaches reduce training cost by controlling Gaussian population, optimizing densification, or simplifying parameter updates.
For instance, LightGaussian~\cite{fan2023lightgaussian} and MiniSplatting~\cite{fang2024mini} use importance-driven pruning or simplification to obtain compact representations.
LP-3DGS~\cite{zhang2024lp} and MaskGaussian~\cite{liu2025maskgaussian} instead incorporate learnable selection mechanisms into Gaussian optimization.
Taming 3DGS~\cite{mallick2024taming} combines budget-constrained densification with efficient gradient computation and parameter updates.
Mini-Splatting2~\cite{fang2024mini2} accelerates convergence through aggressive densification and visibility culling, while DashGaussian~\cite{chen2025dashgaussian} jointly schedules rendering resolution and Gaussian growth.
ContraGS~\cite{durvasula2025contrags} enables training directly on compressed Gaussian representations to reduce memory requirements and data movement.

Unlike methods that primarily control representation size or optimization complexity, \proj incorporates per-tile shared-memory constraints into pruning to increase the fraction of tiles eligible for fused rasterization.

System-level approaches further address the memory and execution demands of large-scale training.
GS-Scale~\cite{lee2026gs} and CLM~\cite{zhao2026clm} leverage host-memory offloading and pipelined execution to reduce GPU memory requirements.
GS-Scale additionally defers optimizer updates for Gaussians with zero gradients.
Grendel~\cite{zhao2025on} distributes Gaussian parameters and rendering workloads across multiple GPUs, exchanging Gaussian attributes and partial gradients between partitions.
Splaxel~\cite{jia2026splaxel} instead uses local rendering and global composition through pixel-level communication.

\proj targets a different combination of opportunities: retaining intermediate data on-chip within each rendering GPU and eliminating exactly zero partial-gradient records from inter-GPU communication.
Its sparse decoder directly aggregates the retained gradients without reconstructing dense communication buffers.

\para{3DGS Accelerators.}
Dedicated architectures exploit the computation and memory-access characteristics of Gaussian rendering and training.
GSCore~\cite{lee2024gscore}, GBU~\cite{ye2025gaussian}, and GCC~\cite{pei2025gcc} introduce specialized hardware for Gaussian rendering.
Lumina~\cite{feng2025lumina} reduces repeated computation through caching, while StreamingGS~\cite{zhang2025streaming} introduces a memory-centric streaming dataflow to reduce DRAM traffic.
VR-Pipe~\cite{lee2025vr} augments GPU graphics hardware for Gaussian rasterization, and GRTX~\cite{lee2026grtx} combines streamlined acceleration structures with hardware-supported traversal reuse for Gaussian ray tracing.
For training, GSArch~\cite{he2025gsarch} addresses memory-system bottlenecks, while ARC~\cite{durvasula2025arc} reduces atomic operation overhead through adaptive warp-level reduction.
Cambricon-GS~\cite{wen2026cambricon} combines Gaussian-pixel hybrid parallelism with tiled SSIM-based pipelining.
\proj instead implements its optimizations on existing GPUs and jointly addresses on-chip data residency, tile-level resource constraints, and distributed gradient communication.

\para{Acceleration of Other 3DGS Applications.}
Application-specific systems exploit additional opportunities in SLAM, virtual reality, and large-scale mobile rendering.
GauSPU~\cite{wu2024gauspu}, REACT3D~\cite{wang2025react}, and RTGS~\cite{li2025rtgs} accelerate 3DGS-based SLAM through specialized processing and reductions in redundant computation.
Splatonic~\cite{huang2026splatonic} combines adaptive sparse pixel sampling and Gaussian-parallel pixel rendering to accelerate tracking and mapping.
For virtual reality, MetaSapiens~\cite{lin2025metasapiens} exploits foveated rendering, while Nebula~\cite{zhu2026nebula} combines cloud-client collaborative rendering with accelerated stereo rasterization.
Hierarchical 3DGS~\cite{kerbl2024hierarchical} and Octree-GS~\cite{ren2024octree} organize Gaussians into multiresolution representations to support efficient rendering at different levels of detail.
Voyager~\cite{liu2025voyager} further targets city-scale Gaussian rendering on resource-constrained mobile devices.
These systems primarily address the latency, memory, and energy constraints of online tracking or interactive rendering.
In contrast, \proj focuses on distributed scene training and reduces data movement both within individual GPUs and across the GPUs participating in gradient aggregation.
\section{Conclusion}
\label{sec:conc}

Distributed training is increasingly important for scaling 3D Gaussian Splatting to large scenes, but its efficiency remains constrained by data movement.
We present \proj, which reduces intra-GPU off-chip traffic and inter-GPU gradient communication via three techniques: fused rasterization, hardware-aware pruning, and zero-gradient elision.
With these techniques, \proj achieves up to 6.1$\times$ training efficiency while maintaining the same reconstruction quality.
Nevertheless, \proj represents one step toward addressing the data-movement challenges of distributed 3DGS training.
More fundamental improvements will require algorithm-architecture co-design that jointly optimizes on-chip storage, execution scheduling, and inter-GPU communication in the future.


\bibliographystyle{ACM-Reference-Format}
\bibliography{reference}

\end{document}